\documentstyle[aps,epsf]{revtex} 

\begin{document}
\draft

\title
{
\LARGE \bf
Singularities of the renormalization group flow for
random elastic manifolds}
\author{{\bf D. A. Gorokhov and  G. Blatter}
\\{\it Theoretische Physik, ETH-H\"onggerberg,
CH-8093 Z\"urich, Switzerland} }
\maketitle
\begin{abstract}
We consider the singularities of the zero 
temperature renormalization group flow 
for random elastic manifolds. 
When starting from small scales, this flow goes through two particular points $l^{*}$ 
and $l_{c}$, where the average value of the random squared potential
$\langle U^{2}\rangle $ turnes negative ($l^{*}$) and where the fourth derivative
of the potential correlator becomes infinite at the origin ($l_{c}$). 
The latter point sets the scale where simple perturbation theory breaks down
as a consequence of the competition between many metastable states.
We show that under physically well defined circumstances $l_{c}<l^{*}$
and thus the apparent renormalization of $\langle U^{2}\rangle$ to negative
values does not take place.

\end{abstract}
\pacs{PACS numbers: 05.20.-y, 11.10.Hi, 74.60.Ge, 75.60.Ch, 82.65.Dp}

Consider an elastic manifold with $d$ internal degrees of freedom
embedded into a $(d+N)$-dimensional space in the presence of
a random potential $U({\bf u}, {\bf r})$.
The free energy of the manifold takes the form
\begin{equation}
{\cal F}[{\bf u}]=
\int
d^{d}{\bf r}
\left [\frac{C}{2}{\left (\frac{\partial{\bf u}}{\partial{\bf r}}\right )}^{2}
+U\left ({\bf u}, {\bf r}\right )\right ],
\label{ham}
\end{equation}
with $C$ the elasticity. The random potential $U({\bf u}, {\bf r})$
is assumed to be gaussian with an isotropic correlator
$\langle U({\bf u},{\bf r})   U({\bf u}^{\prime},{\bf r
}^{\prime}) \rangle=
K(|{\bf u}-{\bf u}^{\prime}|)\delta ({\bf r}-{\bf r}^{\prime})$.
The model Hamiltonian (\ref{ham}) describes a large class
of disordered systems including random magnets, dislocations in metals,
and vortices in superconductors\cite{Halpin}.

In a recent paper\cite{Bucheli}
the functional renormalization group (FRG)
approach has been used in order to calculate the critical force for
the depinning of a $(4+N)$-dimensional elastic manifold in the presence of a
weak random potential. The collective-pinning scale has been
identified with the FRG flow point, where the fourth derivative
of the correlator of the random potential at the origin $K^{(4)}(0)$ becomes infinite.
By induction it is possible to show that all the higher even derivatives
also become singular. However, if we look at the equation for the correlator
itself, it turns out that this equation also exhibits a singularity:
at some length $l^{*}$ the average 
value of the random potential squared $K_{l^{*}}(0)$
becomes negative. This situation is, of course, unphysical.
The goal of this note is to show that the collective-pinning
scale $l_{c}$ is {\it always} smaller than the length $l^{*}$
at which the
average value of the pinning potential squared becomes negative. 
We briefly review the method of the calculation
of the collective-pinning length used in Ref.\onlinecite{Bucheli}
for the $(d+N)$-dimensional problem, determine the two length
$l_{c}$ and $l^{*}$ and show that $l_{c}<l^{*}$ for a physical situation.

The one-loop zero temperature FRG equation for a $(d+N)$-dimensional elastic
manifold can be written in the form\cite{Fisher,Balents}
\begin{equation}
\frac{{\partial K}_{l}({\bf u})}{\partial l}=
\left (4-d-4\zeta\right )K_{l}({\bf u})+\zeta u_{\mu} K^{\mu}_{l}({\bf u})+
I\left [\frac{1}{2}K_{l}^{\mu\rho}({\bf u})
K_{l}^{\mu\rho}({\bf u})-K_{l}^{\mu\rho}({\bf u})K_{l}^{\mu\rho}(0)\right ],
\label{frg}
\end{equation} 
where $\zeta$ is the wandering exponent, 
$I=A_{d}/C^{2}\Lambda^{4-d}$ 
($A_{d}={2\pi^{{d}/{2}}}/{\Gamma ({{d}/{2}})}$ and $\Lambda^{-1}$ is the
short scale cutoff ),
and the upper indices $\mu$ and $\rho$
denote the derivative with respect to the cartesian coordinates
$\mu$ and $\rho$.
Differentiating this equation four times with respect to 
${\bf u}$ and substituting ${\bf u}=0$, we obtain\cite{ref}
\begin{equation}
\frac{\partial K^{(4)}_{l}(0)}{\partial l}=
(4-d)K^{(4)}_{l}(0)+\frac{I(N+8)}{3}{K^{(4)}_{l}}^{2}(0),
\label{der4}
\end{equation}
where $K^{(4)}_{l}(0)={{\partial}^{4}K_{l}}/{\partial u_{\mu}^{4}}|_{{\bf u}=0}$.
Integrating Eq.~(\ref{der4}) we find that
the function $K^{(4)}_{l}(0)$ becomes infinite 
at the scale $l_{c}$ defined by 
\begin{equation}
l_{c}=\frac{1}{4-d}\ln\left (1+\frac{3(4-d)}{I(N+8)K_{0}^{(4)}(0)}\right ).
\label{lc}
\end{equation}
The scale $l_{c}$ defines the collective-pinning radius 
$R_{c}={e^{l_{c}}}/{\Lambda}.$ 

On the other hand, differentiating Eq.~(\ref{frg}) twice with
respect to ${\bf u}$ and setting ${\bf u}=0$ we obtain 
\begin{equation}
\frac{\partial K^{(2)}_{l}(0)}{\partial l}=
(4-d-2\zeta )K^{(2)}_{l}(0).
\label{der2}
\end{equation}  
This equation again can be easily solved and substituting the expression
for $K_{l}^{(2)}(0)$ into Eq.~(\ref{frg}) with ${\bf u}=0$
we find that $K(0)$ becomes negative at the scale\cite{ref1}
\begin{equation}
l_{c}^{*}=\frac{1}{4-d}
\ln \left (1+\frac{2(4-d)K(0)}{IN {K^{(2)}(0)}^{2}}\right ).
\label{lc*}
\end{equation}
The scheme used in Ref.\onlinecite{Bucheli} for the determination of the collective
pinning radius $R_{c}$ is valid only if $l_{c}<l^{*}$.
Thus, we have to prove that indeed in a physical situation
$l^{*}>l_{c}$. Taking
into account Eqs.~(\ref{lc}) and (\ref{lc*}) we can write this
inequality in the form
\begin{equation}
\frac{3N}{2(N+8)}<
\frac{K_{0}(0)K_{0}^{(4)}(0)}{{\left [{K_{0}^{(2)}}(0)\right ]}^{2}}.
\label{ineq}
\end{equation}
Next, let us show that the inequality (\ref{ineq}) is satisfied for 
any physical correlator $K({\bf u})$ with a {\it positive} Fourier transform
$K({\bf k})=\int d{\bf u}K({\bf u}) \exp (i{\bf k}{\bf u })$\cite{ref2} ;
the distribution of the Fourier components of the potential $U$
is then given by the product 
${\cal P}(U_{{\bf k}})\propto\prod_{\bf k}
\exp\left (-{U_{\bf k}^{2}}/{2K ({\bf k})}\right )$ and is well defined
as long as $K({\bf k})>0$.
The condition $K({\bf k})>0$ implies that $K({\bf u})$ can be represented in the
form 
\begin{equation}
K({\bf u})=\int d{\bf u}^{\prime} 
P({\bf u}-{\bf u}^{\prime})P({\bf u}^{\prime}). 
\label{repr}
\end{equation}
Taking into account that
$\Delta K({\bf u}=0)=N K^{(2)}(0)$ and 
$\Delta^{2} K({\bf u}=0)=\left [{N(N+2)}/{3}\right ]K^{(4)}(0)$,
with $\Delta$ the Laplace operator in the $N$-dimensional space,
we can rewrite the right hand side of the inequality (\ref{ineq}) in the form
\begin{equation}
\frac{K_{0}(0)K_{0}^{(4)}(0)}{{\left [{K_{0}^{(2)}}(0)\right ]}^{2}}=
\frac{3N}{N+2}
\frac{\Delta^{2}K({\bf u }=0)K({\bf u}=0)}
{{\left [\Delta K({\bf u}=0)\right ]}^{2}}.
\label{iden}
\end{equation}
Using the Schwarz inequality
$(y_{1},y_{2})\le ||y_{1}||\thinspace ||y_{2}||$ and Eq.~(\ref{repr}),
with the scalar product and the norm defined as 
$(y_{1},y_{2})=\int  d{\bf u} y_{1}({\bf u}) y_{2}({\bf u})$
and $||y_{1}||={\left [\int  d{\bf u}y_{1}^{2}({\bf u}) \right ]}^{{1}/{2}}$
(in particular, 
$K(0)={||P||}^{2}$, $\Delta K (0) = (P, \Delta P)$, and
$\Delta^{2} K (0)={||\Delta P||}^{2}$),
we arrive at the result
\begin{equation}
\frac{K_{0}(0)K_{0}^{(4)}(0)}
{{\left [{K_{0}^{(2)}}(0)\right ]}^{2}}
=
\frac{3N}{N+2}
\frac{\Delta^{2}K({\bf u }=0)K({\bf u}=0)}
{{\left [\Delta K({\bf u}=0)\right ]}^{2}}
\ge \frac{3N}{N+2}.
\label{ineq1}
\end{equation} 
We then can reformulate the condition (\ref{ineq})
to read
\begin{equation}
\frac{3N}{2\left (N+8\right )}\le
\frac{3N}{N+2},
\end{equation}
which is always true and hence
$l^{*}>l_{c}$ for  any {\it physical} correlator $K_{0}({\bf u})$.
At the point $l_{c}$ the third derivative $K_{l_{c}}^{(3)}(0)$
exhibits a jump\cite{Fisher} and the FRG equations (\ref{der4}) and (\ref{der2})
break down as we have used the fact that
all odd derivatives of the correlator vanish at the point ${\bf u}=0$
in their derivation.
The appearance of new terms in the FRG equations will then prevent
the function $K_{l}(0)$ from taking negative values at scales beyond $l_{c}$.


\begin{thebibliography}{99}



\bibitem{Halpin}for a general rewiew see 
 G.~Forgacs, R.~Lipowsky, and 
Th.~M.~Nieuwenhuizen, in ``{\it Phase Transitions and
Critical Phenomena}'', Vol.~{\bf 14}, edited by C.~Domb and
J.~Lebowitz (Academic Press, London, 1991);
G.~Blatter, M.V.~Feigel'man,
V.B.~Geshkenbein, A.I.~Larkin, and V.M.~Vinokur, 
Rev.~Mod.~Phys.~{\bf 66}, 1125 (1994);
T.~Halpin-Healy and Y.-C.~Zhang, 
{\bf 254}, 215 (1995) 
and references therein.
\bibitem{Bucheli}H.~Bucheli, O.~S.~Wagner, V.~B.~Geshkenbein,
A.~I.~Larkin, and G.~Blatter, Phys. Rev. B {\bf 57}, 7642 (1998).
\bibitem{Fisher}D.~S.~Fisher, Phys. Rev. Lett. {\bf 56}, 1964 (1986).
\bibitem{Balents}L.~Balents and D.~S.~Fisher, Phys. Rev. B {\bf 48},
5949 (1993).
\bibitem{ref} We start with the correlator
$K_{0}({\bf u})$ which has zero odd derivatives at the point
${\bf u}=0$. From Eq.~(\ref{frg}) we see that all odd derivatives remain
zero (at least as long as $l<l_{c}$).
\bibitem{ref1}We want to point out that 
the result~(\ref{lc*}) remains true even if $\zeta$ depends on $l$.
\bibitem{ref2}
We wish to point out that there is no proof that $K_{l}({\bf k})$
remains positive for any $l$ under the RG transformation.


\end{thebibliography}
\end{document}